\RequirePackage{fix-cm}
\documentclass[twocolumn,epjc3,showpacs,preprintnumbers,amsmath,amssymb]{svjour3}
\smartqed  % flush right qed marks, e.g. at end of proof
\RequirePackage{graphicx}
\RequirePackage[numbers,sort&compress]{natbib}
\usepackage{graphicx,amsmath,amsfonts,latexsym,amssymb,graphics,epsfig,subfigure,color,makeidx,hyperref}
\usepackage{multirow}
\usepackage{ulem}

\journalname{Eur. Phys. J. C}

\begin{document}

\title{Equatorial orbits of spinning test particle in rotating boson star}

\author{Yu-Peng Zhang\thanksref{e1,addr1}
       \and Yan-Bo Zeng\thanksref{e2,addr1}
       \and Yong-Qiang Wang\thanksref{e3,addr1,addr2}
       \and Shao-Wen Wei\thanksref{e2,addr1}
       \and Yu-Xiao Liu\thanksref{e5,addr1,addr2}
}

\thankstext{e1}{e-mail:zyp@lzu.edu.cn}
\thankstext{e2}{e-mail:zengyb19@lzu.edu.cn}
\thankstext{e3}{e-mail:yqwang@lzu.edu.cn, corresponding author}
\thankstext{e4}{e-mail:weishw@lzu.edu.cn}
\thankstext{e5}{e-mail:liuyx@lzu.edu.cn}

\institute{Lanzhou Center for Theoretical Physics, Key Laboratory of Theoretical Physics of Gansu Province, Lanzhou University, Lanzhou 730000, China \label{addr1}
\and
Institute of Theoretical Physics \& Research Center of Gravitation, Lanzhou University, Lanzhou 730000, China \label{addr2}
\and
School of Physical Science and Technology, Lanzhou University, Lanzhou 730000, China \label{addr3} }

 \maketitle

\begin{abstract}
In this paper, we study the circular orbit of the spinning test particle in the background of a rotating boson star. Using the pole-dipole approximation and neglecting the back-reaction of the spinning test particle on the spacetime, the equation of motion of the spinning test particle is described by the Mathisson-Papapetrou-Dixon equation. We solve this equation under the Tulczyjew spin-supplementary condition and obtain the four-momentum and four-velocity of the spinning test particle. Quite different from the spinless particle, the effective potential of the spinning particle with zero orbital angular momentum goes to infinite at the center of the rotating boson star. This will lead to the fact that the spinning particle can not pass through the center of the boson star. However, when the spin angular momentum and orbital angular momentum satisfy $2\bar{s}+\bar{l}=0$, the effective potential is not divergent anymore and the spinning particle can pass through the center of the rotating boson star. {We still investigate how the spin affects the structure of the circular orbits and we find that the spin will induce the larger or smaller regions of no circular orbits, unstable circular orbits, and stable circular orbits.} Moreover, the radius and energy of the circular orbit will be decreased or increased by the particle spin. These results will have an important application in testing the gravitational waves in the boson star background.\\

\end{abstract}

\maketitle

\section{Introduction}\label{scheme1}

Boson stars are formed by the self-gravitating complex scalar fields {with everywhere} regular localized configurations, and they were firstly introduced in the end of 1960s \cite{Feinblum1968,Kaup1968,Ruffini1969}. They can have a lager mass with a small size without any event horizon and singularity. Since boson stars were proposed in 1968, the research of boson stars has received a lot of attention and various boson stars have been proposed \cite{Lynn:1988rb,Schunck:1996he,Yoshida:1997qf,Mielke1981,Colpi1986,Astefanesei:2003qy,Schunck:2003kk,Liebling:2012fv,Grandclement:2014msa,Meliani:2015zta,Cao:2016zbh,Herdeiro:2017fhv,Alcubierre:2018ahf,Delgado:2020udb,Minamitsuji:2018kof,Guerra:2019srj,Li:2020ffy,Herdeiro:2020kvf,Zeng:2021oez,Dzhunushaliev:2021vwn,Brito:2015pxa,wang2019}. As the mimicker of the black hole \cite{Guzman:2009zz}, {boson stars are important objects in Cosmology} \cite{Liddle1992,Lee1992,Phillippe1992}. The possible gravitational waves emitted from boson stars have been still studied extensively \cite{Balakrishna:2006ru,Palenzuela:2006wp,Sanchis-Gual:2018oui,Bustillo2021} in terms of the nonlinear evolution. Especially the study in Ref.~\cite{Bustillo2021} showed that the {binary Proca boson stars can also be the possible degenerate gravitational wave source of the GW190521 \cite{LIGOScientific:2020iuh}.}

Continued research on boson stars shows that boson stars are the very important test beds of the strong gravity system and the corresponding properties should be investigated in detail. The problem of the equations of motion for a test particle in the background of strong gravity system is critical for understanding the motion of various stars in the galaxy. Note that, a boson star is everywhere regular and there is no phenomenon that a test particle plunges into the event horizon of a black hole. Such properties ensure that novel orbits will be obtained in the backgrounds of boson stars \cite{Grandclement:2014msa,Grandclement:2016eng,Grould12017,Collodel:2017end,Zhang:2021xhp,Delgado:2021jxd}. For example,  a test particle with zero orbital angular momentum can pass through the center of the boson star. Usually, the test particle approach ignores the inner-structure of a small body. For a test particle with a spin, such as a small body with a dipole inner structure, the spin-curvature force will lead the corresponding equation of motion of the test particle will no longer be geodesics. {It was still shown that the test particles can also have the orbits with both zero radial and angular velocities ($u^r=u^\varphi=0$) at some special points \cite{Grandclement:2014msa,philippe2017,Collodel2018}. Therefore, it is natural to ask these questions: can a test particle pass through the center of the boson star when its spin (dipole inner-structure) can not be neglected? Does a spinning test particle have the similar orbits obtained in Refs. \cite{Grandclement:2014msa,philippe2017,Collodel2018}. In this paper, we will consider a spinning test particle under the pole-dipole approximation and use the Mathisson-Papapetrou-Dixon (MPD) equations \cite{Mathisson,Papapetrou1951a,Papapetrou1951c,Dixon} to study how the particle spin affects the corresponding motion in rotating scalar boson stars. We will mainly focus on the circular orbits and the noncircular retrograde orbits with zero velocities peaks.}

This paper is organized as follows. In Sec.~\ref{reviewbsandmpd}, we briefly review the construction of a boson star and obtain the four-momentum and four-velocity of a spinning test particle in the equatorial plane by solving the MPD equations. In Sec.~\ref{effpvandco}, we derive the radial effective potential of the spinning test particle in the background of the rotating boson star and investigate the properties of the motion of the spinning test particle. Finally, the brief conclusion and discussion are given in Sec.~\ref{Conclusion}.

\section{MPD equation solutions in background of a Boson star} {\label{reviewbsandmpd}}

Firstly, we give a brief review about the construction of a boson star solution. A boson star is a strong gravity system formed by the gravitationally bound Bose-Einstein condensate for a light scalar
with long de Broglie wavelength. It is described by the following action
\begin{equation}
S =  \int{d^4 x \sqrt{-g}
  \left[\frac{1}{16\pi G}R - \nabla_\mu\Phi\nabla^\mu\Phi^* - V(\Phi)\right]}.
\label{action}
\end{equation}
We take the Planck units and set the Newton gravitational constant $G$, the speed of light $c$, and the Planck constant $\hbar$ to be unity ($G=c=\hbar=1$). We only consider the mini boson star \cite{Schunck2003} and the potential $V(\Phi)$ is taken as the simplest form
\begin{equation}
V(\Phi)=\mu^2\Phi\Phi^*,
\label{potentialofphi}
\end{equation}
where $\mu$ is the mass parameter of the scalar field $\Phi$.

The geometry of a rotating boson star is stationary and axisymmetric, the corresponding ansatz of the complex scalar field $\Phi$ is defined as follows
\begin{equation}
\Phi=\phi(r,\theta)\exp\left[i(\omega t -k\varphi)\right],
\label{scalarf}
\end{equation}
where the scalar field is dependent on the coordinates of the spacetime $(t, r, \theta, \varphi)$ and the properties of a rotating boson star are controlled by the frequency $\omega$ and angular number $k$. {Varying the action \eqref{action} with metric and scalar field, one can obtain the corresponding field equations as follows
\begin{eqnarray}
R_{\mu\nu}-\frac{1}{2}Rg_{\mu\nu}&=& 8\pi T_{\mu\nu},\label{greq}\\
\nabla_\mu\nabla^\mu \Phi        &=& \mu^2\Phi.\label{eqsf}
\end{eqnarray}
The stationary and axisymmetry of a rotating boson star demand that the metric of rotating boson star can be described by the form \cite{Herdeiro2015} as follows
\begin{eqnarray}
ds^2&=&g_{tt}dt^2+g_{rr}dr^2+g_{\theta\theta}d\theta^2+2g_{t\varphi}dt d\varphi+g_{\varphi\varphi}d\varphi^2\nonumber\\
&=&- e^{2F_0}dt^2 +~e^{2F_2}r^2\sin^2\theta(d\varphi-Wdt)^2\nonumber\\
&&+ e^{2F_1}\left(dr^2+r^2d\theta^2\right),
\label{metric}
\end{eqnarray}
where the metric functions $F_0$, $F_1$, $F_2$, and $W$ are the functions of the coordinate $r$ and $\theta$.

With the help of the metric ansatz \eqref{metric} and the field configuration \eqref{scalarf}, one can numerically obtain the explicit forms of the field equations \eqref{greq} and \eqref{eqsf} \cite{Herdeiro2015}. For the rotating scalar boson star, the metric functions and scalar field should have the reflection symmetry. Thus, we adopt the following boundary conditions at $\theta = \pi/2$:
\begin{eqnarray}
\partial_\theta F_i(r,\frac{\pi}{2})=\partial_\theta W(r,\frac{\pi}{2})=0
\label{boundary0}
\end{eqnarray}
and
\begin{eqnarray}
\left\{
\begin{array}{cc}
\partial_\theta \phi(r,\frac{\pi}{2})=0,~~~\textrm{even parity},\label{ds}\\
~~\phi(r,\frac{\pi}{2})=0,~~~\textrm{odd parity}
\label{Ads}.
\end{array}\right.
\label{boundary1}
\end{eqnarray}
The nonsingular properties of a boson star means that the metric functions and scalar field on the pole points at $\theta=0$ and $\theta=\pi$ satisfy
\begin{eqnarray}
\partial_\theta F_i(r,0) &=& \partial_\theta W(r,0) \,= \phi(r,0) =0, \label{boundary2}\\
\partial_\theta F_i(r,\pi) &=&  \partial_\theta W(r,\pi)= \phi(r,\pi) =0. \label{boundary3}
\end{eqnarray}
The asymptotic flat behavior of a rotating boson star also demands that
\begin{eqnarray}
\lim_{r\to\infty} F_i = \lim_{r\to\infty} W = \lim_{r\to\infty} \phi =0.
\label{boundary4}
\end{eqnarray}

Combining the equations of motion \eqref{greq} and \eqref{eqsf} and the boundary conditions \eqref{boundary0}-\eqref{boundary4}, one can numerically obtain the solutions of the scalar boson stars. In this paper, we use the same method in Ref. \cite{wang2019} to get the numerical solutions of the rotating boson stars. The boson star solutions can also be obtained by using the spectral solver KADATH \cite{philippe2010,Grandclement:2014msa}.} 

Next we start to solve the equations of motion for a spinning test particle in the background of the rotating boson star. For a spinning test particle, the corresponding equations of motion will no longer be geodesics and they are described by the MPD equations \cite{Mathisson,Papapetrou1951a,Papapetrou1951c,Dixon,Wald:1972sz,phdthesis}
\begin{eqnarray}
\frac{D P^{\mu}}{D \lambda} &=& -\frac{1}{2}R^\mu_{\nu\alpha\beta}u^\nu S^{\alpha\beta},\label{equationmotion1}\\
\frac{D S^{\mu\nu}}{D \lambda} &=&P^\mu u^\nu-u^\mu P^\nu, \label{equationmotion2}
\end{eqnarray}
where $P^{\mu}$, $S^{\mu\nu}$, and $u^{\mu}$ are the four-momentum, spin tensor, and four-velocity of the spinning test particle along the trajectory, respectively. However, the MPD equations can not uniquely define the motion of the spinning test particle. Hence an additional spin supplementary condition is necessary \cite{Wald:1972sz,Lukes-Gerakopoulos2014,Costa2017,Georgios2017}. Actually, the spin supplementary condition is related to the center of mass of the spinning test particle with different observers \cite{Wald:1972sz,Lukes-Gerakopoulos2014,Costa2017,Georgios2017}. Recently, the works \cite{Alexei2016,Deriglazov:2018vwa,Deriglazov:2017jub} showed that the expression of the three-acceleration contains the divergent terms as the three-velocity satisfies $v\to c$, where $c$ is the speed of light.  Such divergence can be avoided by adding a non-minimal spin-gravity interaction through gravimagnetic moment \cite{Alexei2016}. In this paper, we ignore above behaviors and take the Tulczyjew spin supplementary condition \cite{Tulczyjew}
\begin{equation}
P_\mu S^{\mu\nu}=0.
\label{supplementarycondition}
\end{equation}
The four-momentum $P^\mu$ of the spinning test particle keeps timelike along the trajectory {\color{red}and it} satisfies
\begin{equation}
P^\mu P_\mu=-m^2\label{normal1},
\end{equation}
where $m$ is the mass of the spinning test particle.

We only consider the equatorial motion of the spinning test particle with spin-aligned or anti-aligned orbits. All the components of the four-momentum and spin tensor should satisfy $P^\theta=0$ and $S^{\theta \mu}=0$. Therefore, the non-vanishing independent components of the equatorial orbits are $P^t$, $P^r$, {\color{red}$P^\varphi$, and $S^{r\varphi}$}. By using the spin-supplementary condition (\ref{supplementarycondition}), we get \cite{Hojman1977}
\begin{eqnarray}
S^{r t}=-S^{r\varphi}\frac{P_{\varphi}}{P_t}, ~~~~S^{\varphi t}=S^{r\varphi}\frac{P_r}{P_t}.
\label{spintensor}
\end{eqnarray}
The conservation of the spin for the spinning test particle demands that
\begin{equation}
s^2=\frac{1}{2}S^{\mu\nu}S_{\mu\nu}=S^{\varphi r}S_{\varphi r}+S^{t r}S_{t r}+S^{t\varphi}S_{t\varphi}.
\label{spin}
\end{equation}
Combining  (\ref{normal1}), (\ref{spintensor}), and (\ref{spin}), we obtain the independent $(r-\varphi)$ component of the spin tensor as follows
\begin{equation}
S^{r\varphi}=\frac{s}{m}\frac{P_{t}}{H},
\end{equation}
where the function $H$ is
\begin{equation}
H=\sqrt{g_{rr}(g_{\varphi t}^2 - g_{\varphi\varphi}g_{tt})}.
\end{equation}
Then the non-vanishing components of the spin tensor in the background of the rotating boson star are
\begin{eqnarray}
S^{r\varphi}&=&-S^{\varphi r}=\bar{s}\frac{P_{t}}{H},\nonumber\\
S^{rt}&=&-S^{tr}=-S^{r\varphi}\frac{P_\varphi}{P_t}=-\bar{s}\frac{P_\varphi}{H},\label{spinnozo}\\
S^{\varphi t}&=&-S^{t\varphi}=S^{r\varphi}\frac{P_r}{P_t}=\bar{s}\frac{P_r}{H},\nonumber
\end{eqnarray}
where the parameter $\bar{s}=\frac{s}{m}$ is the per unit mass spin angular momentum of the test particle and the spin direction is perpendicular to the equatorial plane.

The stationary and axisymmetry mean that there are a timelike Killing vector $\xi^\mu=(\partial_t)^\mu$ and a spacelike Killing vector $\eta^\mu=(\partial_\varphi)^\mu$. Due to the spin-curvature force, the related conserved quantities of the spinning test particle  will be different from the test point particle. For a Killing vector $\mathcal{K}^\mu$, the conserved quantity is \cite{Hojman1977,phdthesis}
\begin{equation}
\mathcal{C}=\mathcal{K}^\mu P_\mu- \frac{1}{2} S^{\mu \nu}\nabla_\nu\mathcal{K}_{\mu}.
\label{Eq:Conserved_quantity}
\end{equation}
In the background of a rotating boson star with the metric (\ref{metric}), there are a timelike Killing vector $\xi^\mu=(\partial_t)^\mu$ and a spacelike Killing vector $\eta^\mu=(\partial_\varphi)^\mu$. The two Killing vectors satisfy
\begin{eqnarray}
S^{\mu \nu}{\xi}_{\mu;\nu}&=&S^{\mu\nu}\xi^\beta\partial_\nu g_{\beta \mu},\\
S^{\mu \nu}{\eta}_{\mu;\nu}&=&S^{\mu\nu}\eta^\beta\partial_\nu g_{\beta \mu}.
\end{eqnarray}
Then we can get the two conserved quantities \cite{Hojman1977}
\begin{eqnarray}
m\bar{e}&=&-\mathcal{C}_t=-\xi^\mu P_\mu+\frac{1}{2} S^{\mu \nu}{\xi}_{\mu;\nu}\nonumber\\
&=&-P_t-\frac{1}{2}\frac{\bar{s}}{H}P_t\partial_rg_{t\varphi}+\frac{1}{2}\frac{\bar{s}}{H}P_\varphi\partial_r g_{tt},\label{conservedenergy}\\
m\bar{j}&=&\mathcal{C}_{\varphi}=\eta^\mu P_\mu-\frac{1}{2} S^{\mu \nu}{\eta}_{\mu;\nu}\nonumber\\
&=&P_\varphi-\frac{1}{2}\frac{\bar{s}}{H}P_\varphi\partial_r g_{\varphi t}+\frac{1}{2}\frac{\bar{s}}{H}P_t\partial_r g_{\varphi\varphi},\label{conservedmomentum}
\end{eqnarray}
where $\bar{e}=\frac{e}{m}$ and $\bar{j}=\frac{j}{m}$. The parameters $e$, $m$, and $j$ are the energy, mass, and total angular momentum of the spinning test particle, respectively. We redefine the total angular momentum as $\bar{j}=\bar{s}+\bar{l}$ in terms of the orbital angular momentum $\bar{l}$ and the spin angular momentum.

Solving Eqs. (\ref{normal1}), (\ref{conservedenergy}), and (\ref{conservedmomentum}), we get the non-vanishing  components of the four-momentum:
\begin{eqnarray}
\!\!\!P_t  &=& -\frac{2\,m\,H \left(2\,\bar{e}\,H - \bar{e}\,\bar{s}\,\partial_r g_{\varphi t} - \bar{j}\,\bar{s}\,\partial_r g_{tt}\right)} {4 H^2 + \bar{s}^2\big[ (\partial_r g_{\varphi t})^2 -\partial_r g_{\varphi\varphi}\partial_r g_{tt}\big]},~~~~~\label{memantumpt}\\
P_\varphi  &=& \frac{2\,m\,H \left(2\,\bar{j}\,H - \bar{e}\,\bar{s}\,\partial_r g_{\varphi \varphi} + \bar{j}\,\bar{s}\,\partial_r g_{\varphi t}\right)}{4H^2+ \bar{s}^2\big[(\partial_r g_{\varphi t})^2 - \partial_r g_{\varphi\varphi}\partial_r g_{tt}\big]},\label{memantumpp}
\end{eqnarray}
and
\begin{equation}
(P^r)^2 =-\frac{m^2+g^{\varphi\varphi}P_\varphi^2+2g^{\phi t}P_\varphi P_t+g^{tt}P_t^2}{g_{rr}}.  \label{memantumpr}
\end{equation}
Due to the trajectories of the test particle are independent of the affine parameter $\lambda$ \cite{Dixon,Georgios2017}, we set the affine parameter $\lambda$ as {the coordinate time} and choose $u^t=1$. We can solve the four-velocity $u^\mu$ by using the equations of motion (\ref{equationmotion1}) and (\ref{equationmotion2})  and the components of $S^{\mu\nu}$ in (\ref{spinnozo}) \cite{Hojman2013,Zhang:2016btg}
\begin{eqnarray}
\frac{DS^{tr}}{D\lambda} \!\!&=&\!\! P^t\dot{r}-P^r=-\frac{\bar{s}}{2H}g_{\varphi \mu}R^\mu_{\nu\alpha\beta}u^\nu S^{\alpha\beta}, \label{spinvelocityequation1}\\
\frac{DS^{t\varphi}}{D\lambda}
\!\! &=&\!\!  P^t\dot{\varphi}-P^\varphi
= \frac{\bar{s}}{2H}g_{r \mu}R^\mu_{\nu\alpha\beta}u^\nu S^{\alpha\beta}.  ~~~~ \label{spinvelocityequation2}
\end{eqnarray}

Finally, we can get the non-vanishing components of the four-velocity as follows
\begin{eqnarray}
\dot{r}&=&\frac{b_1 c_2-a_2c_1}{a_1 a_2 - b_1 b_2},\label{radial_v}\\
\dot{\varphi}&=&\frac{b_2 c_1-a_1c_2}{a_1 a_2-b_1 b_2},\label{angular_v}
\end{eqnarray}
where the functions $a_1$, $b_1$, $c_1$, $a_2$, $b_2$, and $c_2$ are defined as
\begin{eqnarray}
a_1&=&P^t+\frac{\bar{s}}{2H}R_{\varphi r\mu\nu}S^{\mu\nu},\\
b_1&=&\frac{\bar{s}}{2H}R_{\varphi \varphi\mu\nu}S^{\mu\nu},\\
c_1&=&-P^r+\frac{\bar{s}}{2H}R_{\varphi t\mu\nu}S^{\mu\nu},\\
a_2&=&P^t-\frac{\bar{s}}{2H}R_{r \varphi\mu\nu}S^{\mu\nu},\\
b_2&=&-\frac{\bar{s}}{2H}R_{rr \mu\nu}S^{\mu\nu},\\
c_2&=&-P^\varphi-\frac{\bar{s}}{2H}R_{rt\mu\nu}S^{\mu\nu}.
\end{eqnarray}
Then the corresponding orbital frequency parameter $\Omega$ of the test particle is
\begin{equation}
\Omega\equiv\frac{u^{\varphi}}{u^t}=\dot{\varphi}.
\label{angular-frequency}
\end{equation}

\section{Effective potentials and orbits}{\label{effpvandco}}

Using the radial effective potential is an easy way to determine the radial motion of a test particle. For the spinning test particle in the equatorial plane of the rotating boson star, the radial momentum is proportional to the radial velocity. Thus, we decompose the radial momentum (\ref{memantumpr}) as follows
\begin{eqnarray}
    (u^r)^2&=&\left(A \bar{e}^2+B \bar{e}+C\right)\nonumber\\
    &\propto& \left(\bar{e}-\frac{-B+\sqrt{B^2-4AC}}{2A}\right)\nonumber\\
    &\times& \left(\bar{e}-\frac{-B-\sqrt{B^2-4AC}}{2A}\right),
    \label{effectivepotentiala}
\end{eqnarray}
where the functions $A$, $B$, and $C$ are
\begin{eqnarray}
A&=&-\frac{4 m^2 g_{rr}^2}{D} \Bigg\{-\bar{s} \partial_r g_{\varphi\varphi} \bigg[2 g_{\varphi t} \left(2 H-\bar{s} \partial_r g_{\varphi t}\right)\nonumber\\
&+&\bar{s} \partial_r g_{\varphi\varphi} ~g_{tt}\bigg]+4 g_{\varphi \varphi}^2 g_{rr} g_{tt}\nonumber\\
&+&g_{\varphi \varphi} \bigg[\bar{s} \partial_r g_{\varphi t} \left(4 H-\bar{s} \partial_r g_{\varphi t}\right)-4 g_{\varphi t}^2 g_{rr}\bigg]\Bigg\},\\
B&=&\frac{8 \bar{j} m^2 g_{rr}^2}{D} \Bigg\{-g_{\varphi t} \bigg[\bar{s}^2 \left(\partial_r g_{\varphi\varphi} \partial_r g_{tt}+\partial_r g_{\varphi t}^2\right)\nonumber\\
&+&4 g_{\varphi \varphi} g_{rr} g_{tt}\bigg]-\bar{s} \bigg[g_{\varphi \varphi} \partial_r g_{tt} \left(2 H-\bar{s} \partial_r g_{\varphi t}\right)\nonumber\\
&-&\partial_r g_{\varphi\varphi} g_{tt} \left(2 H+\bar{s} \partial_r g_{\varphi t}\right)\bigg]+4 g_{\varphi t}^3 g_{rr}\Bigg\},\\
C&=&-\frac{m^2 g_{rr}}{D} \Bigg\{-8 g_{\varphi t}^2 g_{rr} \bigg[\bar{s}^2 \left(\partial_r g_{\varphi t}^2-\partial_r g_{\varphi\varphi} \partial_r g_{tt}\right)\nonumber\\
&+&2 g_{rr} g_{tt} \left(2 g_{\varphi \varphi}+\bar{j}^2\right)\bigg]+4 g_{\varphi \varphi} g_{rr} \bigg[4 \bar{j}^2 g_{rr} g_{tt}^2 \nonumber\\
&+&\bar{s}^2\left(2 g_{tt} \left(\partial_r g_{\varphi t}^2-\partial_r g_{\varphi\varphi} \partial_r g_{tt}\right)-\bar{j}^2 \partial_r g_{tt}^2\right)\bigg]\nonumber\\
&-&8 \bar{j}^2 \bar{s} g_{\varphi t} g_{rr} \partial_r g_{tt} \left(-\bar{s} \partial_r g_{\varphi t}-2 H\right)+16 g_{\varphi t}^4 g_{rr}^2\nonumber\\
&+&4 \bar{j}^2 \bar{s} \partial_r g_{\varphi t} g_{rr} g_{tt} \left(-\bar{s} \partial_r g_{\varphi t}-4 H\right) +16 g_{\varphi \varphi}^2 g_{rr}^2 g_{tt}^2\nonumber\\
&+&\bar{s}^4 \left(\partial_r g_{\varphi t}^2-\partial_r g_{\varphi\varphi} \partial_r g_{tt}\right)^2\Bigg\},
\end{eqnarray}
where
\begin{equation}
D=\bigg[\bar{s}^2 \left(\partial_r g_{\varphi\varphi} \partial_r g_{tt}-(\partial_r g_{\varphi t})^2\right) + 4 H^2\bigg]^2.
\end{equation}
We can define the effective potential of the spinning test particle with the four-momentum pointing toward future by using the positive square root of Eq. (\ref{effectivepotentiala}) as follows \cite{misnerthornewheeler}
\begin{equation}
V_{\text{eff}}  = \frac{-B+\sqrt{B^2-4AC}}{2A}.
\label{effective-potential-r}
\end{equation}
%Here, the positive square root corresponds to the four-momentum pointing toward future, the negative one
%\begin{equation}
%V_{\text{eff}}' = \frac{-B-\sqrt{B^2-4AC}}{2A}
%\label{effective-potential-r-negative}
%\end{equation}
%corresponds to the past-pointing four-momentum \cite{misnerthornewheeler}.

\subsection{circular orbits}

Now, we have got the four-momentum and four-velocity of the spinning test particle in the equatorial plane of the rotating boson star. In this subsection, we will focus on the properties of the circular orbit $r=r_{\text{co}}$ of the spinning test particle, for which we have
\begin{eqnarray}
\dot{r}|_{r=r_{\text{co}}}     &=&  0,\label{co_energy}\\ % ~\left(\bar{e}=V_{\text{eff}}\right)
\ddot{r}|_{r=r_{\text{co}}}    &=&  0. %~ \left(\frac{dV_{\text{eff}}}{dr}=0\right).
\end{eqnarray}
{The above two equations correspond to 
\begin{equation}
\bar{e}=V_{\text{eff}}|_{r=r_{\text{co}}}
\label{co_condition}
\end{equation} 
and 
\begin{equation}
\frac{dV_{\text{eff}}}{dr}\big|_{r=r_{\text{co}}}=0.
\label{condition1}
\end{equation} 
If we further have 
\begin{equation}
\frac{d^2V_{\text{eff}}}{dr^2}\big|_{r=r_{\text{co}}}>0,
\label{condition2}
\end{equation} 
then we will obtain a stable circular orbit. One can also define the effective potential \cite{Delgado:2021jxd} by using 
\begin{equation}
V_{\text{eff,new}}=(u^r)^2.
\label{new_eff_potential}
\end{equation}
With this new effective potential \eqref{new_eff_potential}, the conditions for the stable circular orbits will be \cite{Delgado:2021jxd}
\begin{equation}
\frac{dV_{\text{eff,new}}}{dr}\big|_{r=r_{\text{co}}}=0
\end{equation}
and
\begin{equation}
\frac{d^2V_{\text{eff,new}}}{dr^2}\big|_{r=r_{\text{co}}}<0.
\label{condition3}
\end{equation}
Although the conditions \eqref{condition2} and \eqref{condition3} for stable circular orbits look different, they are essentially equivalent.} We have obtained the effective potential (\ref{effective-potential-r}) of the spinning test particle with the four-momentum pointing toward future and shown that it is dependent on the spin and angular momenta of the test particle. It can be shown that when the particle spin is zero, the effective potential is the same as that of the point particle \cite{Zhang:2021xhp}.

\begin{figure*}[!htb]
\includegraphics[width=\linewidth]{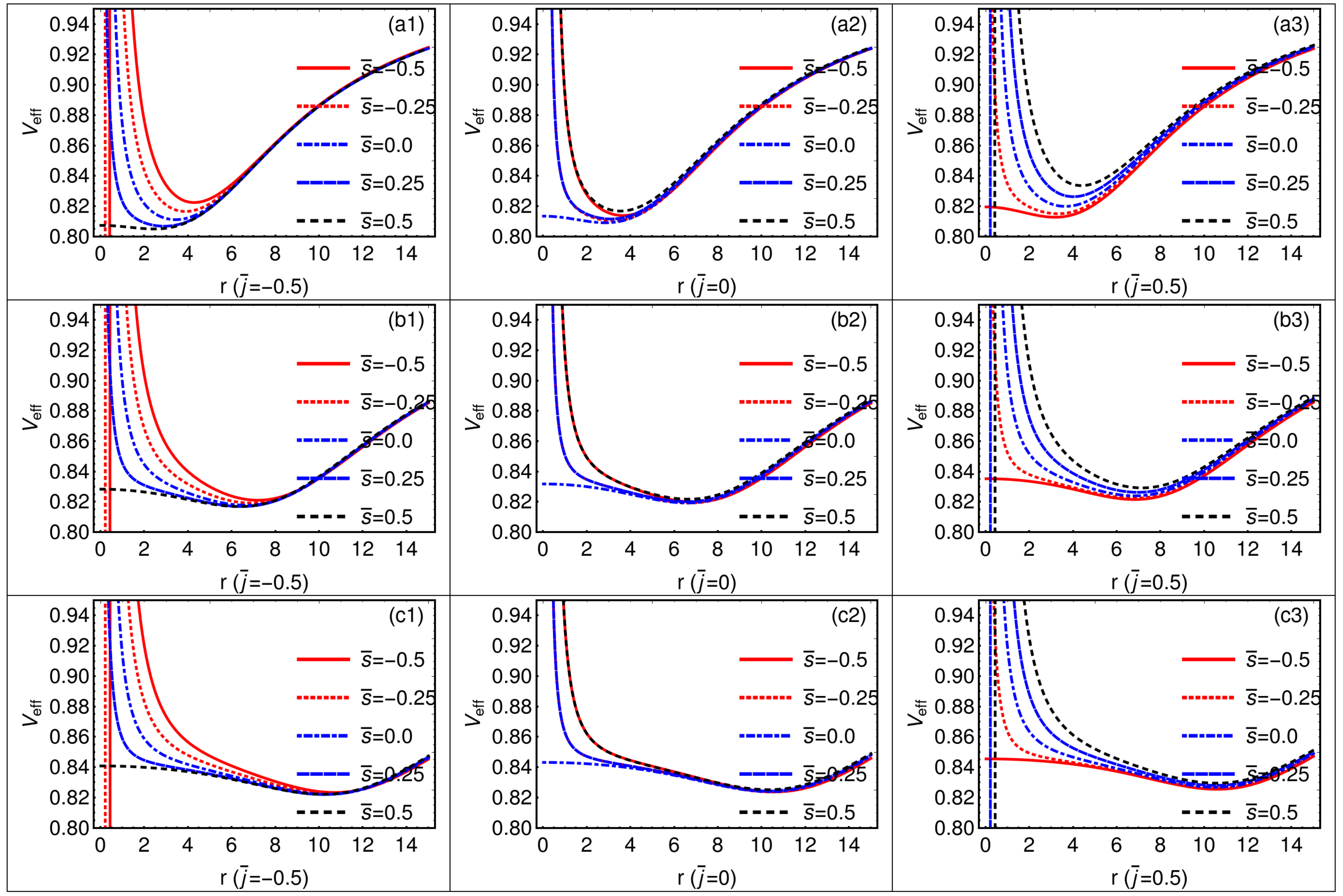}
\caption{The effective potentials of the spinning test particle for different values of the particle spin $\bar{s}$, total angular momentum $\bar{j}$, and angular number $k$. The angular number $k$ is taken as $k=(1, 2,3)$ for subfigures (a1-a3), (b1-b3), (c1-c3), respectively. Other parameters are set as $\omega=0.89$, $\bar{s}=(-0.5,-0.25,0,0.25,0.5)$, and $\bar{j}=(-0.5,0,0.5)$.}
\label{eff_potential_m123}
\end{figure*}

\begin{figure*}[!htb]
	\includegraphics[width=\linewidth]{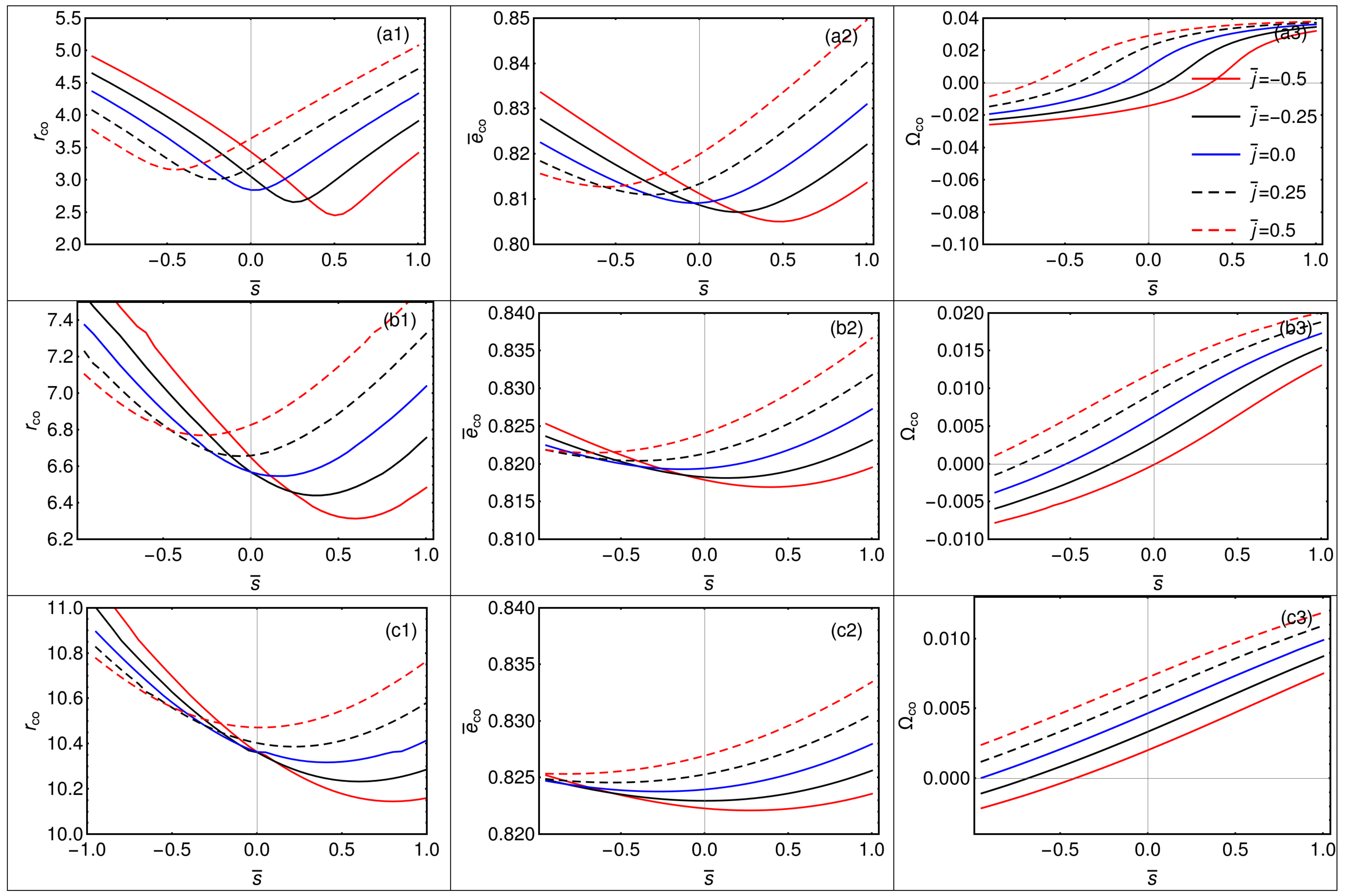}
	\caption{The relations between the circular orbit parameters and the test particle spin. The parameters $\omega$ and $k$ is taken as $\omega=0.89$ and $k=(1, 2,3)$ for subfigures (a1-a3), (b1-b3), (c1-c3), respectively.}
	\label{radius_co_m123}
\end{figure*}

{In this section, we firstly study the properties of stable circular orbits in the backgrounds of rotating boson star with $\omega=0.89$ and $k=(1,2,3)$. We give the effective potentials with different values of the particle spin $\bar{s}$, total angular momentum $\bar{j}$, and angular number $k$ in Fig. \ref{eff_potential_m123}.} We can get the radius of the circular orbit by finding the location of the minimum of the effective potential. The corresponding angular frequency and energy can also be obtained by using Eqs. (\ref{angular-frequency}) and (\ref{co_energy}). Figure \ref{radius_co_m123} shows how the circular orbit parameters (the radius $r_{\text{co}}$, energy $\bar{e}_{\text{co}}$, and angular frequency $\Omega_{\text{co}}$) depend on the test particle spin.

Checking the effective potentials for spinning test particle, we observe some novel behaviors that are not obtained from the point-like test particle. We list them as flollows:

(a) For a spinning particle with zero angular momentum, the spin-curvature force will cause the effective potential at the center of the rotating boson star to be infinite. It means a spinning particle can not pass through the center of the rotating boson star when its orbital angular momentum is zero.

(b) For a spinning particle with fixed orbital angular momentum, when the spin angular momentum changes from an negative value to a positive one, the radius $r_{\text{co}}$ and the particle energy $\bar{e}_{\text{co}}$ for the circular orbit will decrease first and then increase. With the change of the total angular momentum $\bar{j}$ and spin angular momentum $\bar{s}$, the angular frequency for the circular orbit will transform from positive to negative, which will make the test particle move in the opposite angular direction. See the details in Fig. \ref{radius_co_m123}

\begin{figure*}[!htb]
	\includegraphics[width=\linewidth]{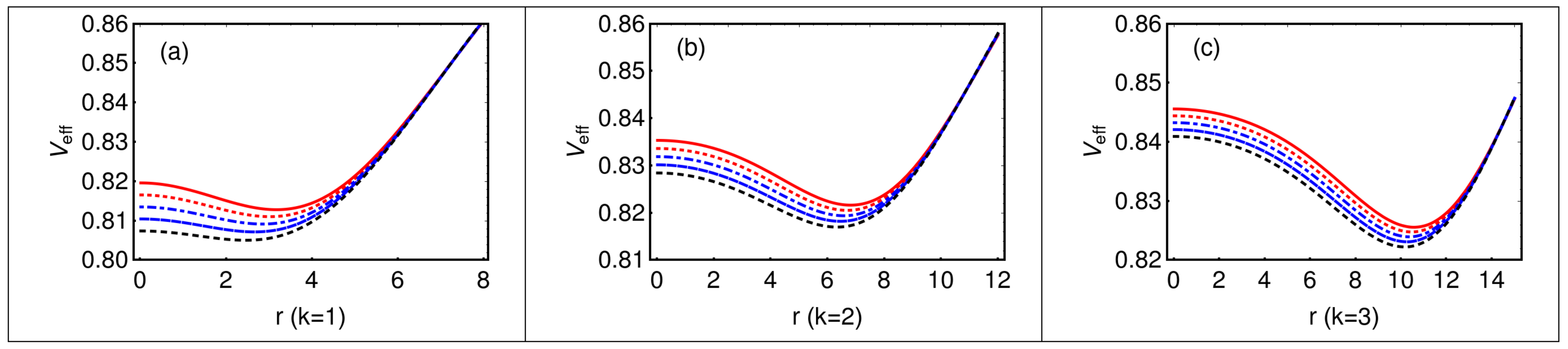}
	\caption{{The effective potentials of the spinning test particle with $\bar{j}+\bar{s}=0$. From top line to bottom line, the corresponding spin and orbital angular {\color{red}momenta} are taken as $(\bar{s}, \bar{j})=(-0.5, 0.5)$, $(-0.25, 0.25)$, $(0, 0)$, $(0.25, -0.25)$, $(0.5, -0.5)$.}
%  $(\bar{s}=-0.5, \bar{j}=0.5), (\bar{s}=-0.25, \bar{j}=0.25), (\bar{s}=0.0, \bar{j}=0.0), (\bar{s}=0.25, \bar{j}=-0.25), (\bar{s}=0.5, \bar{j}=-0.5)$.
  }
	\label{eff_l0_m123}
\end{figure*}

(c) The most surprising result is that when the total angular momentum and spin angular momentum of the spinning particle satisfy $\bar{s}+\bar{j}=0$, the effective potential at the origin is still finite. That is to say, when the spin angular momentum and the orbital angular momentum satisfy $2\bar{s}+\bar{l}=0$, a spinning particle with suitable energy could pass through the center of the rotating boson star. See the details in Fig. \ref{eff_l0_m123}.

\begin{figure}[!htb]
	\includegraphics[width=0.8\linewidth]{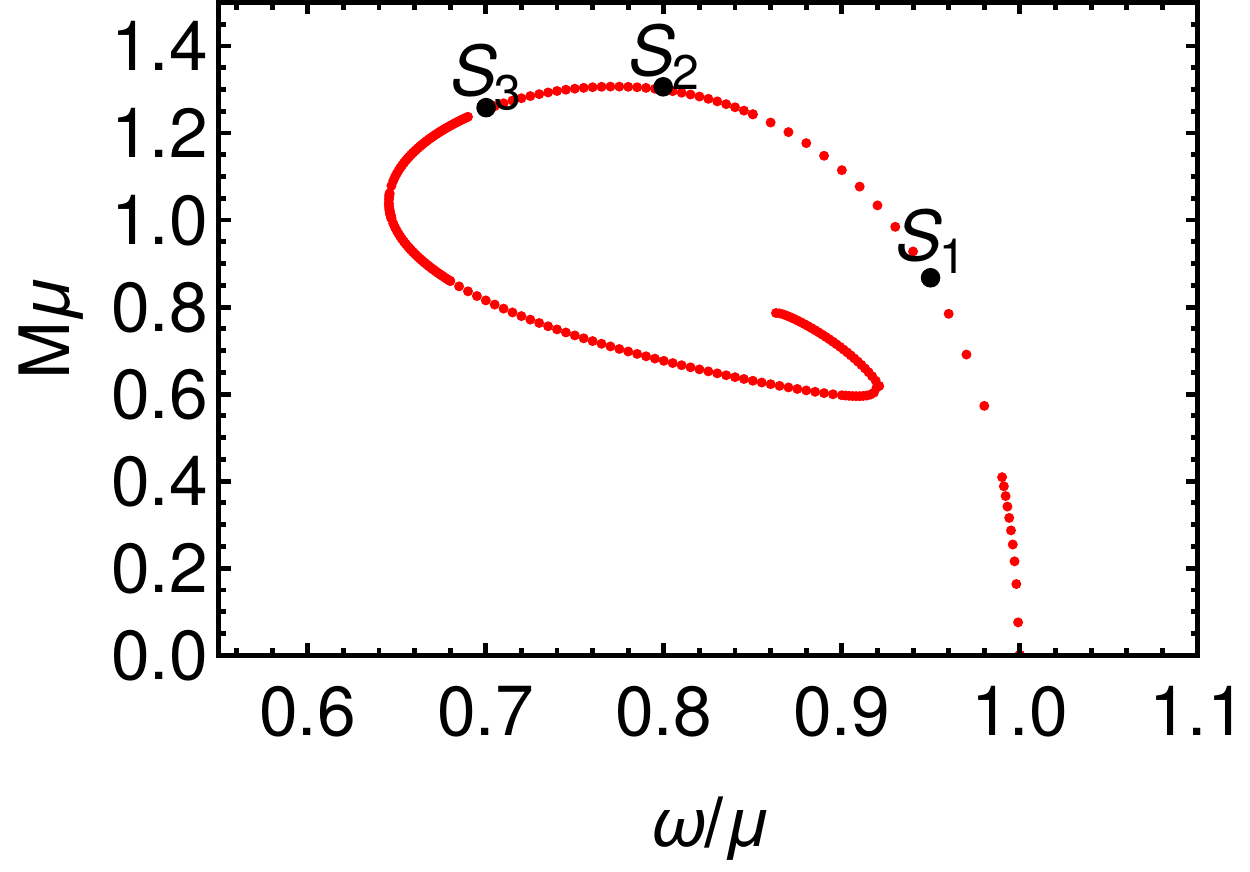}
	\caption{Plots of the ADM mass $M$ versus the frequency $\omega$. Here we set the parameter $k=1$.}
	\label{frequency}
\end{figure}

\begin{figure*}[!htb]
	\includegraphics[width=\linewidth]{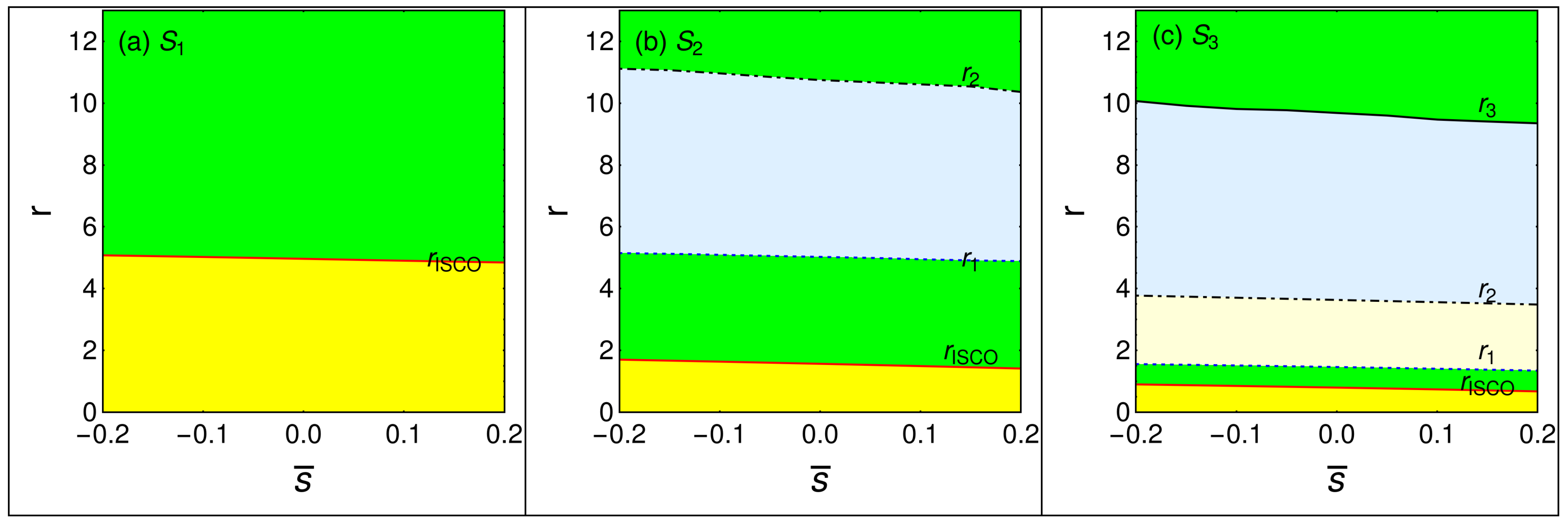}
	\caption{Structures of circular orbits of spinning test particle in the scalar boson star solutions $S_1$, $S_2$, and $S_3$. Here, there are no circular orbits in the yellow region,  there are stable circular orbits in the green region, there are unstable circular orbits in lightblue region.  
	}
	\label{orbits_structure}
\end{figure*}

{It has been shown that the properties of boson star depend on the angular number $k$ and the frequency $\omega$. For a fixed angular number $k$, a rotating boson star will transform from the low rotating state to the highly relativistic rapidly rotating state with the change of the frequency $\omega$. In Refs. \cite{Grandclement:2014msa,Meliani:2015zta}, the authors showed that the rotating boson star will possess a circular orbit with the smallest radius $r_{\text{ICO}}$ (innermost circular orbit) and all the circular orbits (i.e. $r>r_{\text{ICO}}$) are found in rotating boson star are stable. However, when the orbits of the test particles transform from the corotating states to the counterrotating states, not all the circular orbits are stable. References \cite{Cao:2016zbh,Delgado:2021jxd} have studied the structure of the equatorial timelike circular orbits for a point-like test particle in rotating scalar boson stars. These two works showed that in a rotating scalar boson star, there is a region without any circular orbits with radius $r<r_{\text{ISCO}}$, and there is a region that only has unstable circular orbits. 
	
In this paper, we only consider three scalar boson star solutions with $k=1$ and $\omega=(0.95, 0.80, 0.70)$ to study how the spin affects the structure of the circular orbits. For the three rotating scalar boson stars that we considered (see the details in Fig. \ref{frequency} and Table \ref{mass_points}), they could have different structure of circular orbits. The solution $S_1$ has an ISCO and which is the same for corotating and counterrotating orbits. The solution $S_2$ has the stable orbits for radius $r$ between $r_{\text{ISCO}}$ and $r_1$, as well as for $r>r_2$, and the orbits with $r$ between $r_1$ and $r_2$ are unstable. For the solution $S_3$, the counterrotating orbits with $r$ between $r_{\text{ISCO}}$ and $r_1$ are stable, there are no circular orbits (stable or unstable) with $r$ between $r_1$ and $r_2$, there are unstable circular orbits with $r$ between $r_2$ and $r_3$, and stable orbits with $r$ for $r>r_3$.

For the spinning test particle, such structure of the circular orbits will be different. To show how the spin affects the structure of the circular orbits, we consider the spin is in the range of $\bar{s}\in(-0.2,0.2)$ and obtain the dependence of the radii $(r_{\text{ISCO}}, r_1, r_2, r_3)$ on the spin $\bar{s}$. We give the results in Fig. \ref{orbits_structure}. We find that the spin will lead the $r_{\text{ISCO}}$ decrease or increase, and the region without any circular orbits is still existing. The regions for the stable and unstable circular orbits are still existing and the particle spin will induce the corresponding boundary radii decrease or increase.}

\begin{table}[!htb]
	\begin{center}%\renewcommand\arraystretch{1.4}
	\begin{tabular}{ c c c c c c c }
	\hline
	model&$\omega$&$\frac{M}{m_p^2/\mu}$&$r_{\text{ISCO}}$&$r_1$&$r_2$&$r_3$\\
	\hline
	$S_1$ & 0.95  & 0.864         & 4.970    &   -     &  -      &-        \\
	$S_2$ & 0.80  & 1.308         & 1.574    &   5.035 & 10.779  & -       \\
	$S_3$ & 0.70  & 1.261         & 0.803    &   1.470 & 3.640   & 9.680   \\
	\hline
    \end{tabular}
	\caption{Structures of circular orbits for a point-like test particle in the scalar boson star solutions $S_1$, $S_2$, and $S_3$.}
	\label{mass_points}
	\end{center}
\end{table}

\subsection{orbits with zero velocities peaks}

{It has been shown that the point-like test particle could possess the orbits with both zero radial and zero angular velocities at some special points, i.e., $u^r=u^\varphi=0$ \cite{Grandclement:2014msa,philippe2017,Collodel2018}. With the help of the angular effective potential besides the radial effective potential \eqref{effective-potential-r}, one can accurately obtain the orbits parameters. Here, the angular effective potential of the point-like particle is defined as
\begin{equation}
\bar{e}-V_{\text{eff},\varphi}=\bar{e}+\frac{ \bar{J} g_{tt}} {g_{t\varphi}}=0,
\end{equation}
where
\begin{equation}
V_{\text{eff},\varphi}=-\frac{ \bar{J} g_{tt}} {g_{t\varphi}}.
\end{equation}
When the energy of the test particle satisfies $\bar{e}=V_{\text{eff},\varphi}=V_{\text{eff}}$, the test particle could be initially rest \cite{Grandclement:2014msa} or always rest at a special position of the rotating boson star~\cite{Collodel2018}. In this part, we will investigate the properties of such orbits for the spinning test particle. 

Due to the complexity of the angular velocity \eqref{angular_v} of a spinning test particle, it is difficult to get the corresponding formula of the angular effective potential. However, we can firstly specify the fixed total angular momentum $\bar{j}$ and spin angular momentum $\bar{s}$, and then numerically solve the two equations $\dot{r}=0$ and $\dot{\varphi}=0$ to obtain the corresponding energy $\bar{e}$ and radius $r_{\text{static}}$ of the point with zero radial and zero angular velocities. 

To show how the spin angular momentum affects the orbits with both zero radial and zero angular velocities at some special points, we only consider the background of the rotating boson star with $\omega=0.89$ and $k=1$. We choose a fixed orbital angular momentum $\bar{l}=-0.2$ and derive the orbits with varied spin angular momentum $\bar{s}=(-0.1, 0.0, 0.1)$ in Fig. \ref{eff_angular}. For the fixed orbital angular momentum, we find that the radii of the zero-velocity points on the orbit decrease with the spin angular momentum changes. Moreover, when the spin angular momentum increases, the total angular momentum $\bar{j}$ also changes, and the trend of the orbital change induced by the change of the total angular momentum is the same as that induced by the change of the orbital angular momentum of the spinless particle.

\begin{figure*}[!htb]
	\includegraphics[width=\linewidth]{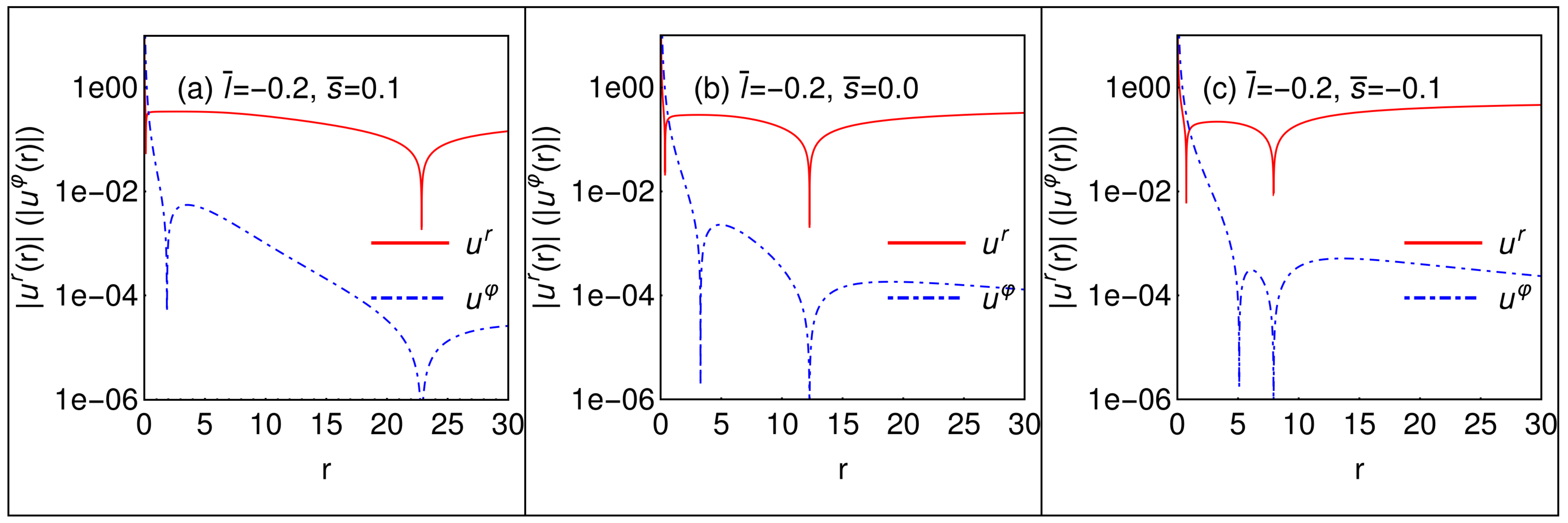}
	\includegraphics[width=\linewidth]{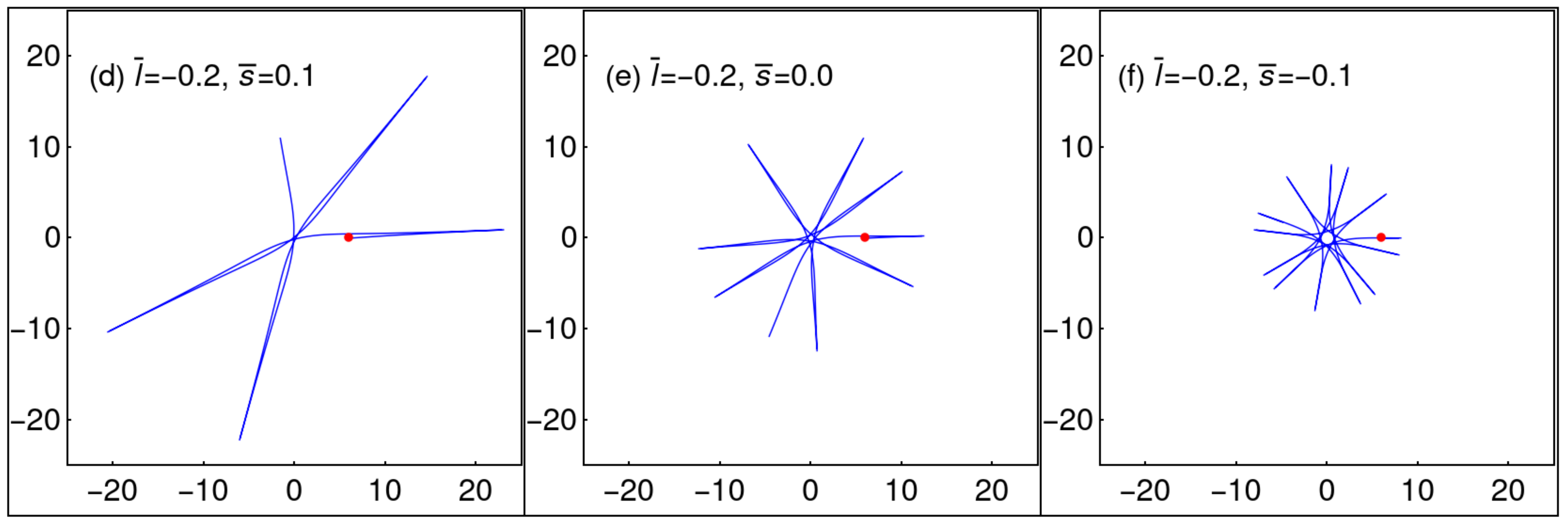}
	\includegraphics[width=\linewidth]{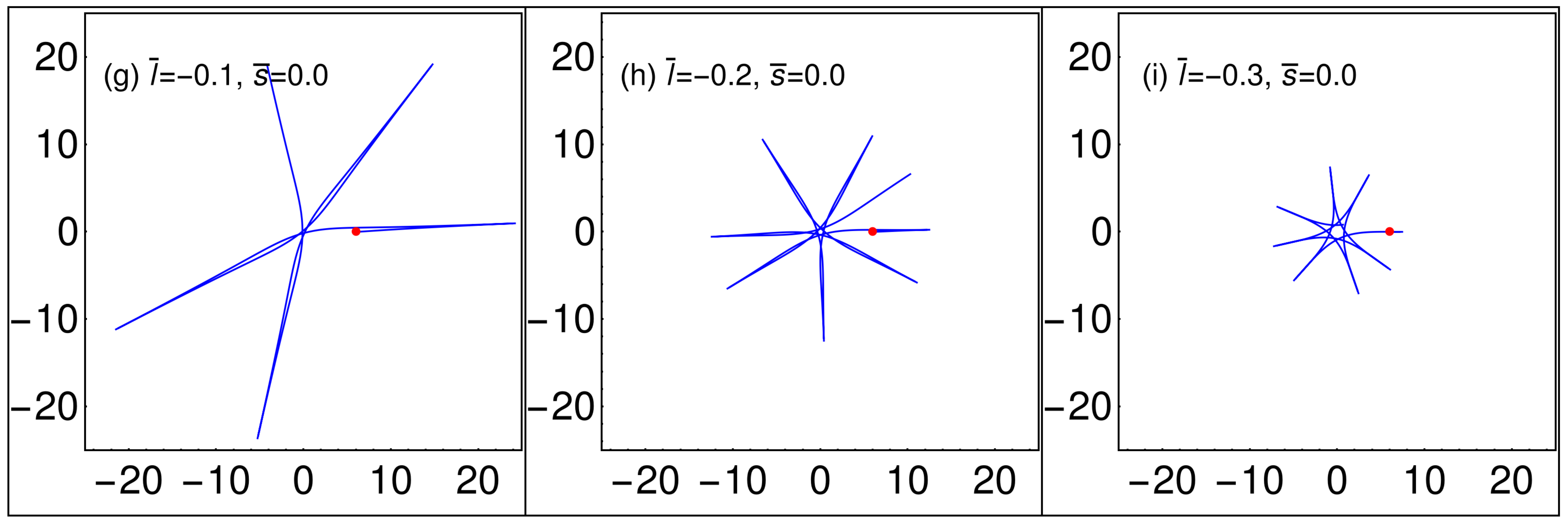}
	\caption{Plots of the radial and angular velocities (up channel) and orbits with zero radial and zero angular velocities. The middle channel describes the orbits of spinning test particle with the same orbital angular momentum and different spin angular momentum. The down channel describes the the orbits of test particle with the different orbital angular momentum and vanishing spin angular momentum.}
	\label{eff_angular}
\end{figure*}

 }

\section{Conclusions and discussion}\label{Conclusion}

In this paper, we investigated the circular orbits of a spinning test particle in the equatorial plane of the rotating boson star with the angular number $k=(1, 2, 3)$ and the frequency $\omega=0.89$. We solved the four-velocity of the spinning test particle and derived the corresponding radial effective potential. We obtained the four-momentum and four-velocity of the spinning test particle by solving the MPD equations in the rotating boson star background. The radial effective potential was obtained in terms of the radial component of the four-momentum $P^r$ for the spinning test particle with the four-momentum pointing toward future.

We found that the particle spin leads to the divergence of the effective potential at the center of the rotating boson star for the test particle with zero orbital angular momentum. We studied the relations between the circular orbit parameters and the particle spin, and showed that the particle spin can increase or decrease the radius and energy of the circular orbit. There is a novel feature for the radial effective potential of the spinning test particle, i.e., when the spin and the total angular momentum of the test particle satisfy $\bar{s}+\bar{j}=0$, the effective potential at the origin is still finite and the test particle can pass through the center of the rotating boson star. {We also investigated how the particle spin affects the structures of the circular orbits in the rotating scalar boson stars. The orbits with static points of spinning test particle were still investigated. Compared with the spinless particle, we found that the spinning test particle has the similar structures of the circular orbits but different boundary radii, and the particle spin will induce the regions for the no circular orbits, stable circular orbits, and unstable circular orbits increase or decrease.} These results will lead to some novel orbits and have an important application in testing the gravitational waves in the boson star background, which will be studied in the near future.

\section{Acknowledgments}

This work was supported in part by the National Key Research and Development Program of China (Grant No. 2020YFC2201503), the National Natural Science Foundation of China (Grants No. 12105126, No. 11875151, No. 12075103, and No. 12047501), the China Postdoctoral Science Foundation (Grant No. 2021M701531), the 111 Project under (Grant No. B20063), the Fundamental Research Funds for the Central Universities (Grant No. lzujbky-2021-pd08), ``Lanzhou City's scientific research funding subsidy to Lanzhou University''.

%\section*{References}

\end{document}